\documentclass[aps,prb,twocolumn,amssymb,color,pdflatex]{revtex4}

\usepackage{graphicx}% Include figure files
\usepackage{graphics}
\usepackage{dcolumn}% Align table columns on decimal point
\usepackage{bm}% bold math
\usepackage{amsmath}
\usepackage{color}

\usepackage{amsmath}

\begin{document}

\newcommand{\red}{\color{red}}
\newcommand{\old}{\color{black}}
\newcommand{\bk}{{\bf k}}
\newcommand{\bp}{{\bf p}}
\newcommand{\bv}{{\bf v}}
\newcommand{\bq}{{\bf q}}
\newcommand{\tbq}{\tilde{\bf q}}
\newcommand{\tq}{\tilde{q}}
\newcommand{\bQ}{{\bf Q}}
\newcommand{\br}{{\bf r}}
\newcommand{\bR}{{\bf R}}
\newcommand{\bB}{{\bf B}}
\newcommand{\bA}{{\bf A}}
\newcommand{\ba}{{\bf a}}
\newcommand{\bE}{{\bf E}}
\newcommand{\bj}{{\bf j}}
\newcommand{\bK}{{\bf K}}
\newcommand{\cS}{{\cal S}}
\newcommand{\vd}{{v_\Delta}}
\newcommand{\tr}{{\rm Tr}}
\newcommand{\kslash}{\not\!k}
\newcommand{\qslash}{\not\!q}
\newcommand{\pslash}{\not\!p}
\newcommand{\rslash}{\not\!r}
\newcommand{\bs}{{\bar\sigma}}
\newcommand{\omt}{\tilde{\omega}}

\newcommand{\qperp}{q_{\perp}}
\newcommand{\qpar}{q_{\parallel}}
\newcommand{\beq}{\begin{equation}}
\newcommand{\eeq}{\end{equation}}

\newcommand{\redtext}[1]{\textcolor{red}{#1}}
\newcommand{\bluetext}[1]{\textcolor{blue}{#1}}
\newcommand{\brown}[1]{\textcolor{brown}{#1}}
\def\intt{\int\limits}
\newcommand{\ket}[1]{| #1 \rangle}
\newcommand{\bra}[1]{\langle #1 |}
\newcommand{\dirac}[2]{\langle #1 | #2 \rangle}
\def\be{\begin{equation}}
\def\ee{\end{equation}}

\title{Semiclassical approach to 2d impurity bound states in Dirac systems}

\author{Kun W. Kim$^1$, T. Pereg-Barnea$^{2}$ and G. Refael$^1$}
\affiliation{$^1$Department of Physics,
California Institute of Technology, 1200 E. California Blvd, MC114-36,
Pasadena, CA 91125 }
\affiliation{$^2$Department of Physics and Centre for Physics of Materials, McGill University, Montreal, Quebec, Canada H3A 2T8}
\date{\today}
\begin{abstract}
The goal of this paper is to provide an intuitive and useful tool for analyzing the impurity bound state problem.  We develop a semiclassical approach and apply it to an impurity in two dimensional systems with parabolic or Dirac like bands.  Our method consists of reducing a higher dimensional problem into a sum of one dimensional ones using the two dimensional Green functions as a guide.  We then analyze the one dimensional effective systems in the spirit of the wave function matching method as in the standard 1d quantum model. We demonstrate our method on two dimensional models with parabolic and Dirac-like dispersion, with the later specifically relevant to topological insulators.
\end{abstract}
\maketitle

\section{Introduction}
The presence of disorder is often considered as a nuisance that degrades the quality of samples and obscures the behavior of clean physical systems. Even in small amounts, however, impurities may induce new phases which are interesting in their own right, and do not have a clean-system analog. Prominent examples are the metal-insulator transition induced by random on site potential\cite{and1958,and1978,abr2010,eve2008}, Cooper-pair breaking transition in  conventional s-wave superconductors by magnetic\cite{and1959,and1961,abr1958,abr1959} and non-magnetic\cite{mar1963, lar1965,ma1985} impurities, impurity-induced spin quantum-hall effect\cite{li2009,gro2009,guo2010}, and the Kondo effect\cite{kon1964,hew1997}.

Understanding the single impurity problem often provides strong intuition for the behavior of a disordered system with a finite impurity density. Using this as motivation, we study the problem of bound states of a single narrow impurity in a variety of host systems. For a narrow impurity, bound states could be found most straightforwardly by solving the Schr\"odinger equation outside and inside the impurity-affected region, and matching the wave functions at the boundary. In 1d, with a delta-function impurity potential, this is particularly simple.  It is also quite straight forward when dealing with two dimensional systems and a one dimensional perturbation such as an edge.  However, for point-like impurities in two-dimensions, a more complicated consideration is required.

Nevertheless, in this work we show that a 2d system with a point-like impurity could be reduced to a 1d problem on a straight trajectory, in which wavefunction matching can be applied. Furthermore, in spatially anisotropic systems (namely, lacking rotational symmetry about the impurity), we show that using a small number of incoming and outgoing beams, straight trajectories allows a remarkably accurate estimate of bound state energies.

The semiclassical approach has been instrumental in providing insightful physical pictures in terms of classical trajectories in complicated quantum systems, especially when impurity scatterers or confining potentials are involved. Examples of such applications include quasiparticle states near extended scatterers in d-wave superconductor\cite{ada1999}, bound states in multidimensional systems with Fermi resonance\cite{noi1979}, the low-energy spectrum of charge carriers in graphene\cite{sto2012}, and Berry phase in graphene\cite{car2008}. The standard semiclassical methods which map complicated multidimensional quantum problems onto 1d quantum problem on simple classical trajectories, however, are approximations, and suffer from limitations that need to be addressed with more sophisticated methods \cite{sto2012, kno1976}.

In the following sections we derive the mapping from 2d to an effective 1d impurity Hamiltonian, and then use it to  find  the bound state energies in several examples of increasing complexity. This mapping should be thought of as a semi-classical description of the 2d impurity problem, where the bound-state energies are obtained by considering a small number of classical incoming and outgoing beams. The mapping from 2d to 1d relies on the Green function of the clean system, which indicates which 'classical' paths are necessary. Our method also approximates the bulk Hamiltonian by its form in the vicinity of minima in momentum space, and assuming a parabolic or Dirac-like dispersion.  We find that with the introduction of appropriate cutoffs using the Pauli-Villars regulators, this approximation remains relatively precise.

The organization of the manuscript is as follows. In Sec.~II we show how the 2d impurity problem can be reduced to a 1d problem for an isotropic Hamiltonian system. In Sec.~III, the method is extended to an anisotropic Hamiltonian and then in Sec.~IV to a band structure with multiple minima. While the extension of our method to any odd dimension is straight forward, the extension to even dimensions is not. In Sec.~V we show how to generalize our method to all even dimensional systems.

\section{wave-matching for single isotropic minimum continuum band}\label{section:isotropic}
\begin{figure}
\includegraphics[width=85mm]{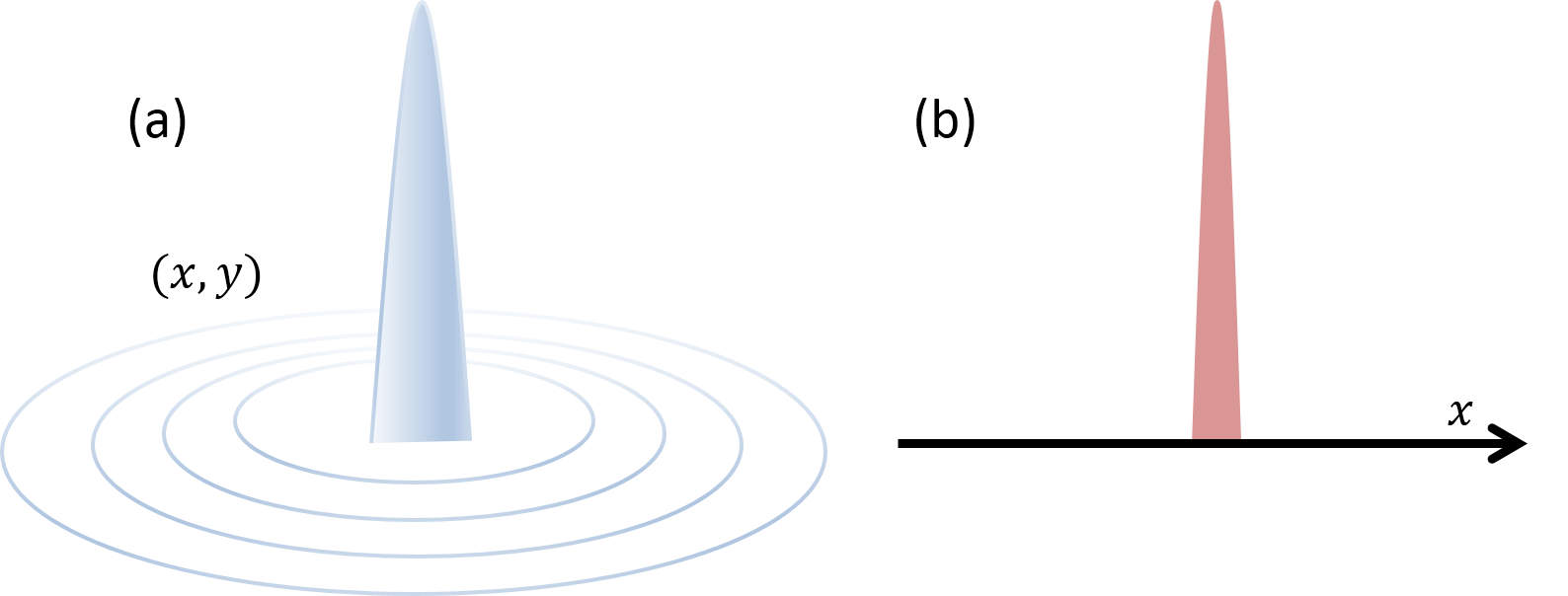}
\caption{ 2d system with a point-like impurity (a) is reduced to the 1d system with a modified impurity strength. Thus, a simple picture of wavefunction matching can be applied to 2d system to obtain a bound state energy associated with a single impurity in 2d. }
\label{fig:motivation}
\end{figure}

\subsection{Review of the one dimensional problem}
Our goal in this section is to derive a method equivalent to the 1d wavefunction matching technique, for finding the bound states of an impurity in an isotropic 2d system. For this purpose, let us briefly review how a bound state associated with a single point-impurity is obtained in a 1d system. The Schr\"odinger equation in this case is simply:
\begin{eqnarray}
[E-H(\partial_x)]\psi (x) = \alpha \delta (x) \psi (x),
\label{scho}
\end{eqnarray}
where  Hamiltonian H is responsible for kinetic part only, and  the impurity scattering strength is $\alpha$.  The Green function of the operator $E-H( \partial_x)$ is given by:
\begin{eqnarray}
G(x)&=&  \intt^\infty_{-\infty} \frac{dk}{2\pi} \frac{e^{ikx}}{E-\tilde{H}(k)},
\label{1dsol}
\end{eqnarray}
where $\tilde{H}(k)$ is the Hamiltonian in momentum representation.  The Green function $G(x)$ is understood as the amplitude of the propagator at $x$, originating from the source at $x=0$.  Eq.\eqref{1dsol} can be computed using contour integration around the upper (lower) half of complex k-plane for positive (negative) x. If the energy, $E$, is within the band, the (real-axis) poles of the integrand each correspond to a plane wave, and the combination of these waves makes up the Green function. If the energy is outside the band, as would be the case for a bound state, the poles are not on the real axis and therefore the wave function decays exponentially with distance. We can still think of such a Green function as a combination of plane waves, but with complex wave vectors.

To obtain a bound state energy associated with the impurity, we could take two paths. Formally, we use the fact that the Green function is the solution of Eq.\eqref{scho} omitting $\alpha\psi(x)$ on the right side of the equality. Therefore:
\be
\psi(x)= G(x)\alpha \psi(0),
\label{g0}
\ee
and we obtain  for a scalar Hamiltonian :
\be
\frac{1}{\alpha}=G(0)=\intt^\infty_{-\infty} \frac{dk}{2\pi} \frac{1}{E-\tilde{H}(k)},
\label{18}
\ee
consistent with the T-matrix formulation\cite{bal2006}.
 
 For a Hamiltonian with internal structure such as sublattice or spin, the Green function and the impurity potential are matrices.  The impurity potential matrix ${\bf \alpha}$, may not be invertible. In this case, for $\vec{\psi} (0)$ to have a non-trivial solution at $x=0$, the following condition is required:
\be
\rm{Det}[I - G(0){\bf \alpha}]=0.
\label{det}
\ee
For simplicity of presentation, in most of this manuscript we consider only scalar problems.  In Sec.~IV, along with extending our method to the case of host systems with multiple low-gap valleys, we also assume a multi-component wave function.

An alternative to the above method, is to solve the 1d equation as Eq.\eqref{scho} simply by matching a freely propagating  plane-wave solution at $x>0$ with a different  plane-wave solution at $x<0$. Integrating the Schr\"odinger equation over the impurity position gives:
\begin{eqnarray}
\alpha\psi(0) - \int_{-\epsilon}^{\epsilon} (E-H) \psi (x) dx=0 \label{1drelation}.
\end{eqnarray}
Indeed,  the same bound energy relation as Eq.\eqref{18} or Eq.\eqref{det} is obtained from this wave function-matching approach by inserting Eq.\eqref{g0} to the above equation.

\subsection{Effective one dimensional problem}
Could we use the same notion of 'wave-function matching' in the context of a bound state in 2d?
Let us start with 2d Schr\"odinger equation:
\begin{eqnarray}
[E-H_{2d}(\vec{r}) ]\Psi (\vec{r}) = \alpha \delta^2 (\vec{r}) \Psi (\vec{r}).
\end{eqnarray}
To solve for a bound state at energy $E$, we can still use Eq.\eqref{g0}, adapted to 2d with
\begin{eqnarray}
G(x) &=&  \int^\infty_{-\infty} \frac{d^2k}{(2\pi)^2} \frac{e^{ikx}}{E-\tilde{H}(k)},
\label{2dgf}
\end{eqnarray}
and:
\begin{eqnarray}
\frac{1}{\alpha}=G(0)= \int \frac{d^2\vec{k}}{(2\pi)^2} \frac{1}{E-\tilde{H}_{2d}(\vec{k})}, \label{2dgeneral}
\end{eqnarray}
where $\tilde{H}_{2d}(\vec{k})$ is the 2d Hamiltonian in momentum representation. For this section, we assume an isotropic Hamiltonian  $\tilde{H}_{2d}(\vec{k})=\tilde{H}_{2d}(k)$. Hence we can write Eq.\eqref{2dgeneral} as:
\begin{eqnarray}
\frac{1}{\alpha}=G(0)= \int \frac{d\theta}{2\pi}\intt_0^{\infty} \frac{k dk}{2\pi} \frac{1}{E-\tilde{H}_{2d}(k)} \label{2dgeneral-k}.
\end{eqnarray}

Eq.\eqref{2dgeneral-k} can not be simply interpreted in terms of  plane  wave matching as its 1d counterpart since contour integration over a complex $k$ can not be used: the integration range is $ 0 \le k<\infty$. Nevertheless, we can proceed using the Kramers-Kronig relation along with the symmetry of the imaginary part of the integrand and obtain an expression analogous to Eq.\eqref{18}.

The Kramers-Kronig relation connects the imaginary and real parts of a complex function, $f(s)$ which is analytical in the upper half plane, $\rm{Im}(s)>0$, and falls off faster than $1/|s|$. To apply it here, we define:
\be
F(s)=\intt_0^{\infty} \frac{k dk}{2\pi} \frac{e^{iks}}{E-\tilde{H}_{2d}(k)}.
\ee
Note that this function is not the Green function in real space, since there is no angular dependence taken into account in the exponent. However, $F(0)=G(0)$. Furthermore, we see that only $\rm{Re}[G(0)]$ plays a role in determining the bound-state energy of an impurity state. We use the Kramers-Kronig relation to write:
\be
\rm{Re}[F(0)]=\intt_{-\infty}^{\infty}\frac{ds'}{\pi s'}\rm{Im}[F(s')].
\ee
This is helpful since the imaginary part of $F(s)$ for $\rm{Im}(s)=0$ obeys:
\be
\rm{Im} [F(s)]=\frac{1}{2}\intt_{-\infty}^{\infty} \frac{k dk}{2\pi} \frac{e^{iks}}{E-\tilde{H}_{2d}(k)},
\ee
with the $k$ integral now stretching over the entire real axis.  It is assumed that the Hamiltonian $\tilde{H}_{2d}(k)$ is an even function of k. 

After these steps we can rewrite the bound state energy condition as:
\begin{eqnarray}
\frac{1}{\alpha}
&=&  \frac{1}{\pi} \int_{-\infty}^{\infty} \frac{ds}{s} Im\left[ \int_0^{\infty} \frac{k dk}{2\pi} \frac{e^{iks}}{E-\tilde{H}_{2d}(k)} \right], \\
&=&  \frac{1}{\pi} \int_{0}^{\infty} \frac{ds}{is} \left[ \int_{-\infty}^{\infty} \frac{k dk}{2\pi} \frac{e^{iks}}{E-\tilde{H}_{2d}(k)} \right].
\end{eqnarray}
Finally, we also eliminate the factor of $k$ in the integrand using a derivative with respect to s, to yield our final expression:
\begin{eqnarray}
\frac{1}{\alpha}&=&  \frac{-1}{\pi} \int_{0}^{\infty} \frac{ds}{s} \frac{\partial}{\partial s} \left[ \int_{-\infty}^{\infty} \frac{dk}{2\pi} \frac{e^{iks}}{E-\tilde{H}_{2d}(k)} \right].
\label{2drelation}
\end{eqnarray}
 The term in the square brackets in Eq.\eqref{2drelation} is completely analogous to the expression used to find the bound state energy of a 1-d solution of a point-impurity, Eq.\eqref{1dsol} and we therefore define:
\begin{eqnarray}
G_{1d} (s) &=& \int_{-\infty}^{\infty} \frac{dk}{2\pi} \frac{e^{iks}}{E-\tilde{H}_{2d}(k)},
\label{2dsolution}
\end{eqnarray}
which can be viewed as the Green function of a 1-d Hamiltonian $H_{1d}(s) \equiv \tilde{H}_{2d}( k \rightarrow  \frac{\partial}{i \partial s})$ in real space. Qualitatively, the effective real-space 1-d Hamiltonian, $H_{1d}$, describes a single direction of the 2-d momentum-space Hamiltonian $\tilde{H}_{2d}$.

The final result of this reasoning is that the bound-state energy $E$ for a 2d point impurity, can be obtained by solving a 1d impurity problem, with a modified potential.  Provided that $G_{1d}(x)$ is the solution of the 1-d Hamiltonian $H_{1d}(x)$, therefore the wave function $\psi_{1d}(x) =G_{1d}(x)$ up to an overall normalization factor for a scalar Hamiltonian. From Eq.\eqref{2drelation} and Eq.\eqref{2dsolution} the effective 1-d Schr\"odinger equation can be written with a modified impurity potential:

\begin{eqnarray}
\left[E-H_{1d}(x) \right]\psi_{1d} (x) =  \delta (x)\alpha'\psi_{1d}(0),
\label{1dsch}
\end{eqnarray}
with $\alpha'$ given by:
\be
\alpha'= -\alpha\left[ \frac{1}{\pi}\int_{0}^{\infty} \frac{ds}{ s} \frac{\partial \psi_{1d} (s)}{\partial s} \right]\frac{1}{\psi_{1d}(0)}.\label{1dredux}
\ee
and $\psi_{1d}(s)$  on the right hand side of Eq.\eqref{1dredux}   emerges as the solution of the effective 1d problem, but does not appear to have any physical significance. 
%Eq.~(\ref{1dredux}) constitute our main result.

\subsection{Example: Free particle}

The first example for our method is the free particle with quadratic
dispersion. We demonstrate the method by finding the 1d effective
Hamiltonian in real space, and solving it in the spirit of the
wavefunction matching method. The free particle example in 2d is a bit
pathological, however, since a high energy cutoff is required, and the bound state energies of point
impurities depend on it. This is seen by inspecting
Eq.\eqref{2dgeneral}: the momentum integration diverges, unless a
cutoff is imposed. This introduces a technical challenge for our
quest to use momentum integration over the entire real-axis. We resolve
it by introducing the cutoff using the Pauli-Villars regularization
technique\cite{zee2010}.

Starting with the free Hamiltonian  and an impurity potential,
\begin{eqnarray}
H_{2d}&=& -\frac{\nabla^2}{2m},\\
 V_{imp}&=& \alpha \delta^{(2)}(\vec{x}).
\end{eqnarray}
We write the Green-function condition for an impurity bound state, and
include the Pauli-Villars regulators $W^2/(k^2+W^2)$:
\begin{eqnarray}
\frac{1}{\alpha} &=&  \int \frac{d^2\vec{k}}{(2\pi)^2} \frac{1}{E-\tilde{H}_{2d}(k)} \left( \frac{W^2}{W^2+k^2} \right).
\end{eqnarray}
According to our recipe, this is equivalent to finding bound states
of the following 1-d effective Hamiltonian:
\begin{eqnarray}
\tilde{H}_{1d}(k) = \frac{k^2}{2m} - \left( E-\frac{k^2}{2m} \right) \frac{k^2}{W^2}.
\label{efffp}
\end{eqnarray}

It is interesting to consider the effective 1d Hamiltonian,
Eq.\eqref{efffp} and the effect that the Pauli-Villars regulator
has. $\tilde{H}_{1d}$ preserves the original dispersion relation, and
a $k^2/2m$ pole is reflected in the Green function. In addition, the
Green function acquires an additional pair of poles at $k_p = \pm i
W $.

\begin{figure}
\includegraphics[width=85mm]{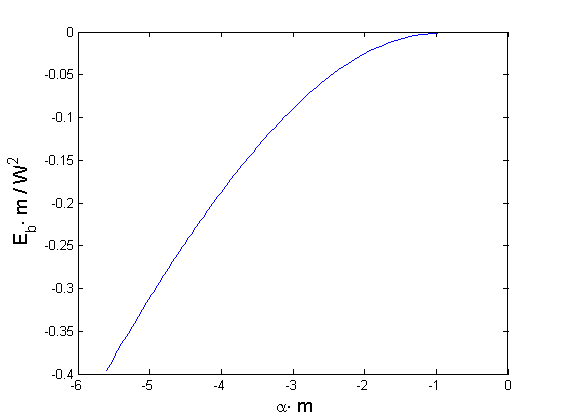}
\caption{ Bound energy $E_b$ of the free particle model with mass $m$ and cutoff $W$ for the regularization is plotted as a function of impurity strength $\alpha$. The bound energy decreases as the impurity strength negatively increases. In the limit of zero impurity strength, the bound energy converges to zero. }
\label{fig:free}
\end{figure}

Now we can treat the problem just like a 1-d problem with a plane
wave solution, but with  multiple plane  waves corresponding to all solutions
of $E\psi_{1d}=\tilde{H}_{1d}\psi_{1d}$ (which correspond to the
poles of the Green function). The solution must satisfy continuity
conditions at $x=0$ of the derivatives:
\begin{eqnarray}
\left[ \frac{\partial^m \psi (x)}{\partial x^m} \right]_{0-}^{0+} = 0,
\end{eqnarray}
for $m=0,1,2$, which provides enough conditions to fix the weights of
the plane waves up to an overall factor on $x<0$ and $x>0$:
\begin{eqnarray}
\psi(x>0) &=& \frac{1}{\lambda_1}e^{-\lambda_1 x} - \frac{1}{\lambda_2}e^{-\lambda_2 x}, \label{psolution1} \\
\psi(x<0) &=& \frac{1}{\lambda_1}e^{\lambda_1 x} - \frac{1}{\lambda_2}e^{\lambda_2 x}. \label{psolution2}
\end{eqnarray}
The exponents are $\lambda_1 = \sqrt{-2mE}$, and $\lambda_2 = W$,
which are the poles of Green function. Now we use
 Eq.\eqref{1dsch}  to find the bound state energy for a given
impurity potential strength. First, we carry out the integration that
yields the effective $\alpha'$  using Eq.\eqref{1dredux}:
\begin{eqnarray}
\alpha' = \frac{\alpha}{\pi} \frac{\lambda_1\lambda_2}{\lambda_2-\lambda_1} \log \left( \frac{\lambda_2}{\lambda_1} \right).
\end{eqnarray}
With this modified impurity potential and the 1d effective Hamiltonian Eq.\eqref{efffp} in real space, one can solve the 1d problem using and get:
\begin{eqnarray}
\frac{1}{\alpha}
&=& \frac{-m/\pi}{1+2mE/W^2} \log\left( \frac{W}{\sqrt{-2mE}} \right).
\label{freerelation}
\end{eqnarray}
As seen from Eq.\eqref{freerelation}, the bound state energy is renormalized by the cutoff and mass, but its qualitative behavior as a function of the impurity potential strength is unchanged. This is depicted in Fig.~\ref{fig:free}.

\subsection{A particle in a Mexican hat band}

Let us study one more isotropic Hamiltonian - a Mexican hat shaped energy band:
\begin{eqnarray}
\tilde{H}_{2d}(k) = J(k^2-k_0^2)^2,
\end{eqnarray}
where J is a constant with units of length cubed. The band covers all
positive energies and therefore the energy of any bound state should be
negative.

\begin{figure}
\includegraphics[width=85mm]{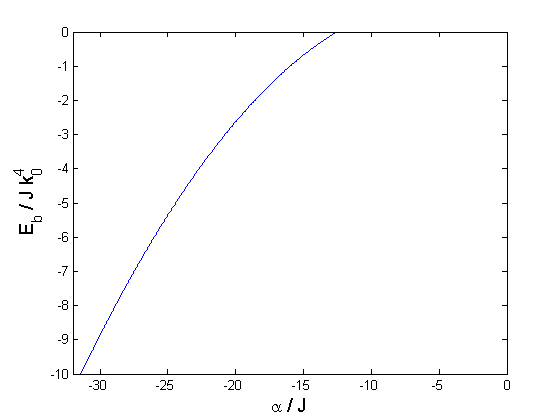}
\caption{ Bound energy $E_b$ of the Mexican hat model is plotted as a function of impurity strength $\alpha$. The general behavior is similar to that of the free particle model except that the bound state appears when $\alpha < -1/4\pi J k_0^2$.  }
\label{fig:mexican}
\end{figure}

The Mexican hat Hamiltonian does not have a UV divergence problems as
the free-particle Hamiltonian, and therefore does not require a
Pauli-Villars regulator. Thus $\tilde{H}_{1d}(k)=\tilde{H}_{2d}(k)$ in
this case.  Incidentally, it also has the same order
of derivatives as the regularized free particle Hamiltonian,
Eq.\eqref{efffp}. Therefore, it obeys the same number of continuity
relations and  we may simply use the same functional form ansatz as in Eq.\eqref{psolution1}-\eqref{psolution2}, with the wave numbers are given by:
\begin{eqnarray}
k_1 &=& i \lambda_1 = k_0 \left( 1+i \sqrt{\frac{-E}{Jk_0^4}} \right)^{1/2}, \\
k_2 &=& i \lambda_2 = k_0 \left( 1-i \sqrt{\frac{-E}{Jk_0^4}} \right)^{1/2},
\end{eqnarray}
which are the result of setting $E=\tilde{H}_{1d}(k)$. Carrying out
the same steps as in the free-particle case, we obtain the relation between the bound state energy and the impurity strength is:
\begin{eqnarray}
\frac{1}{\alpha} &=& \frac{1}{2\pi J}\frac{\log(\lambda_2/\lambda_1)}{\lambda_2^2-\lambda_1^2}, \\
&=& \frac{-1}{4\pi J}\frac{\tan^{-1} \sqrt{-E/Jk_0^4}}{\sqrt{-E/J}},
\end{eqnarray}
where it follows from the above expression that the bound state exists only for attractive potential.  The result is plotted in Fig.~\ref{fig:mexican} with renormalized axes. 

\section{Wavefunction matching for Anisotropic
  bands}\label{section:anisotropic}

When the impurity problem is anisotropic, we can no longer solve for
the bound-state energies exactly using only a small number of incoming
and outgoing beams. Instead, we can still consider simple 1d
wave-functions for the radial part of the problem, and then consider a
superposition over all angles. Here we extend the wavefunction
matching method to such anisotropic systems.

We begin by writing the Hamiltonian in polar momentum coordinates:
$\tilde{H}_{2d}(\vec{k})=\tilde{H}^\theta_{2d}(k)$, with $\theta$
being the momentum direction. Now, when going from Eq.\eqref{2dgeneral} to Eq.\eqref{2dgeneral-k}, the angular integration must be kept:
\begin{eqnarray}
\frac{1}{\alpha}
&=&  \int \frac{d\theta}{2\pi} \int_0^{\infty} \frac{k dk}{2\pi} \frac{1}{E-\tilde{H}^\theta_{2d}(k)}.
\end{eqnarray}
Following  exactly the same steps, we introduce a 1-d wave function along momentum angle $\theta$:
\begin{eqnarray}
G^{\theta}_{1d} (s) &=& \int_{-\infty}^{\infty} \frac{dk}{2\pi} \frac{e^{iks}}{E-\tilde{H}^\theta_{2d}(k)}, \label{e1dsolution}
\end{eqnarray}
which is the Green function of 1-d Hamiltonian  $H^\theta_{1d}(s) \equiv \tilde{H}^\theta_{2d}( k \rightarrow  \frac{\partial}{i \partial s})$. The modified relation between potential strength and associated bound state energy is:

\begin{eqnarray}
\frac{1}{\alpha} =\int \frac{d\theta}{2\pi} \frac{1}{\alpha^\theta},
\label{athet}
\end{eqnarray}
where $1/\alpha^\theta$, just as in the right hand side of  Eq.\eqref{2drelation} , is:
\begin{eqnarray}
\frac{1}{\alpha^{\theta}}=-\frac{1}{\pi}\int_{0}^{\infty} \frac{ds}{ s} \frac{\partial G^{\theta}_{1d} (s)}{\partial s} . \label{aniso}
\end{eqnarray}
 As pointed out above, the Green function $G^{\theta}_{1d}(s)$ is constructed using plane wave solutions of  $\tilde{H}^\theta_{2d}(\frac{\partial }{i\partial s})$ , treated as a 1d Hamiltonian. For a general anisotropic system, the above prescription requires infinitely many directions of 1-d solutions.

A more direct formulation of the $\alpha_{\theta}$ in terms of a 1d Schr\"odinger equation for each $\theta$ direction is as follows. We construct for each $\theta$ a solution of the 1d bound state equation:

\begin{eqnarray}
\left[ E- \tilde{H}_{\theta} \left(\frac{\partial}{i\partial s}\right)   \right]
\psi^{\theta}(s)=\alpha'^{\theta}\delta (s) \psi^{\theta}(s),
\end{eqnarray}
and find $\alpha'^{\theta}$ in terms of $E$ by requiring that the 1d problem has a bound state at energy $E$.
The integrand in the right-hand-side of Eq.\eqref{athet}, $\alpha^{\theta}$, is given in terms  $\alpha'^{\theta}$ and the impurity-state wave function, $\psi^{\theta}(s)$, as:
\begin{eqnarray}
\frac{1}{\alpha^{\theta}}
&=& -\frac{1}{\alpha'^{\theta}} \frac{1}{\pi}\int_{0}^{\infty} \frac{ds}{ s} \frac{\partial \psi^{\theta}(s)}{\partial s}  \frac{1}{\psi^{\theta}(0)}\label{generalrelation1}.
\end{eqnarray}

\subsection{Anisotropic mass}

As a simple example, we consider impurity states in a band described by a parabolic dispersion with an anisotropic mass:
\begin{eqnarray}
\tilde{H}(\vec{k}) &=& \frac{k_x^2}{2m_x}+\frac{k_y^2}{2m_y} \equiv {k^2 \over 2m(\theta)},
\end{eqnarray}
where
\begin{eqnarray}
{1\over m(\theta )} = \frac{cos^2\theta}{m_x}+\frac{sin^2\theta}{m_y}. 
\end{eqnarray}
This problem can be simply solved by rescaling $x$ relative to $y$, and obtaining the isotropic solution. Therefore this calculation should just be a demonstration of using our method. For more complicated band structures this will not be possible (e.g., bands with multiple minima, as discussed below).

\begin{figure}
\includegraphics[width=85mm]{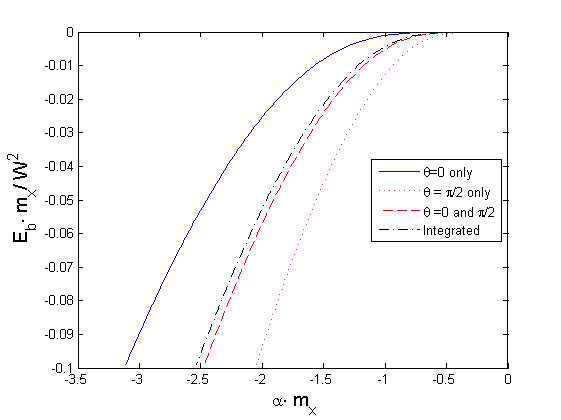}
\caption{ Bound energy associated with a single impurity for two-mass anisotropic band model ($m_y/m_x=2$). The exact bound state energy is obtained by integrating over all momentum angle $\theta$, while taking a straight "1d" path with a certain momentum angle yields slightly deviated bound state energy. Nevertheless, the consideration of only two momentum angle $\theta=0$ and $\pi/2$ gives a fair agreement with the exact result.    }
\label{fig:aniso}
\end{figure}

Let us now construct $G^{\theta}_{1d}$ and $\alpha^{\theta}$. For a given direction we can make use of the relation we obtained before for a free particle, using $m(\theta)$ instead of $m$. Using Eq.\eqref{freerelation}, with $\alpha\rightarrow \alpha^{\theta}$, we obtain:
\begin{eqnarray}\label{eq:theta}
\frac{1}{\alpha^{\theta}}
&=& \frac{-m(\theta)/\pi}{1+2m(\theta)E/W^2} \log\left( \frac{W}{\sqrt{-2m(\theta)E}} \right).
\end{eqnarray}
The last step in this consideration should be the integration of Eq.\eqref{eq:theta} over all momentum directions.  However, we find that in this case it is enough to consider a pair of perpendicular directions - the extrema of $\alpha^{\theta}$ which occur at $\theta=0,\,\pi$ and $\theta=\pi/2,\,3\pi/2$ .  This is shown in Fig.~\ref{fig:aniso} where the dash-dot (black) line shows the result the integration over all directions and the other lines show two different directions (solid, dotted) and their combination (dashed).  As evident from the graph, the combination of two perpendicular directions is very similar to the integrated expression. This is equivalent to evaluating the $\theta$ integral using a discrete sum, which could be used to reduce the number of beams necessary in a 2d problem to a finite and small number.

\section{A band with multiple minima}\label{section:multiple}

Perhaps the most interesting application of our method is to dispersion relations which contain several gap minima. The qualitative picture of a bound state consisting of a simple superposition of several 1d beams each coming from the vicinity of one particular gap minimum is rather intuitive and appealing. Our method allows making this picture quantitative even for complicated band structures.

So far we have identified the single minimum of the band and used an approximated version of the Hamiltonian around the minimum. Then, according to Eq.\eqref{2dgeneral}, the Green function is obtained and integrated with an appropriate cutoff $W$ introduced through the Pauli Villars regulator. The generalization to more than one band minimum, is dividing the integration in Eq.\eqref{2dgeneral} into the summation of multiple integrations with different local Hamiltonians, which are the expansions of the Hamiltonian around each minimum. In many cases where multiple valleys need to be considered, the wavefunction is a multidimensional spinor. Therefore, we need to represent the impurity strength ${\bf \alpha}$ by a matrix and use the determinant condition for bound states, Eq.\eqref{det}.

\begin{eqnarray}
\int \frac{d^2\vec{k}}{(2\pi)^2} \frac{1}{E-\tilde{H}_{2d}(\vec{k})} \approx  \sum_l \int \frac{d^2\vec{k_l}}{(2\pi)^2} \frac{1}{E-\tilde{H}_{l}(\vec{k_l})}.
\end{eqnarray}

Where $l$ goes over the band minima and $H_l$ is the expansion of the Hamiltonian around the $l$th minimum with the appropriate Pauli-Villars regularization included. 
The origin of each $\vec{k_l}$ is set to the center of the minima such that the 'valley' Hamiltonian $\tilde{H}_l(\vec{k_l})$ has maximum symmetry.
Thus, the condition for a bound state is:
\be
\det\left(1-\sum_l \int \frac{d^2\vec{k_l}}{(2\pi)^2} \frac{1}{E-\tilde{H}_{l}(\vec{k_l})}{\bf \alpha}\right)=0.
\label{det1}
\ee

Applying the Kramers-Kronig trick here as well, as in the anisotropic construction, we write the potential strength as a sum:
\begin{eqnarray}
\det\left(1-\sum_l \int \frac{d\theta}{2\pi} \frac{1}{\alpha_l^\theta} {\bf \alpha}\right)=0,\label{generalrelation}
\end{eqnarray}
where $\left(\alpha_l^\theta\right)^{-1}$ is a matrix, given by:
\begin{eqnarray}
\frac{1}{\alpha_l^\theta} =-\frac{1}{\pi}\int_{0}^{\infty} \frac{ds}{ s} \frac{\partial G^{\theta}_{l} (s)}{\partial s}, \label{atmv}
\end{eqnarray}
with
\[
G^{\theta}_l(s)=\int_{-\infty}^{\infty}\frac{dk}{2\pi}\frac{e^{iks}}{E-\tilde{H}_{l}^{\theta}(k)},
\]
the Green function for the Hamiltonian $\tilde{H}_l^{\theta}=\tilde{H}_l(k\hat{x}\cos\theta+k\hat{y}\sin\theta)$.

As described in Sec.~\ref{section:anisotropic}, Eq.\eqref{atmv} can also be interpreted in terms of individual 1d plane waves in each direction and valley. This is made a bit more complicated by taking into account a spinor index. Denoting the spinor indices with  $\sigma,\,\sigma'$, we have
\be
 \left(\alpha_l^{\theta}\right)^{-1}_{\sigma \sigma'}=-\frac{1}{b_{l}^{\theta,\,\sigma'} }  \frac{1}{\pi}\int_{0}^{\infty} \frac{ds}{ s} \frac{\partial [\psi_l^{\theta,\,\sigma '}]_{\sigma}(s)}{\partial s} \frac{1}{[\psi_l^{\theta,\,\sigma' }]_{\sigma'}(0)},
\ee
where $b_{l}^{\theta,\,\sigma'}$ and $[\psi_l^{\theta,\,\sigma'}] _{\sigma}(s)$ are obtained by solving the 1d impurity problem of a particle with the Hamiltonian  $\tilde{H}_{l}^{\theta}(k \rightarrow \frac{\partial}{i\partial s})$ , which is the original 2d Hamiltonian for a particular momentum direction, $\theta$, and with momentum in the vicinity of the bottom of valley $l$. Let us clarify these symbols further, and state the impurity problem  that  needs to be solved:
\be
 [ (E-\tilde{H}_{l}^{\theta}) \psi_l^{\theta,\sigma '}]\, _{\sigma}(s)=\delta_{\sigma\,\sigma'}\delta(s)b_{l}^{\theta,\,\sigma'}[\psi_l^{\theta,\,\sigma'}]_{\sigma}(s).\old
\ee
The index $\sigma'$ indicates to which component the impurity couples, and $b_{l}^{\theta,\,\sigma'}$ is the impurity strength required to induce an impurity state with energy $E$ in the specified valley, direction, and component of the spinor involved. Accordingly, $[\psi_l^{\theta,\,\sigma'}]_{\sigma}(s)$, is the $\sigma$ component at point $s$ of the (valley $l$ and momentum angle $\theta$) wave function of a bound state of an impurity that couples to the $\sigma'$ component.
Note that the impurity strength is multiplied by the $\sigma'$ component of the wave function. 

\subsection{Kane-Mele model}

\begin{figure}
\includegraphics[width=85mm]{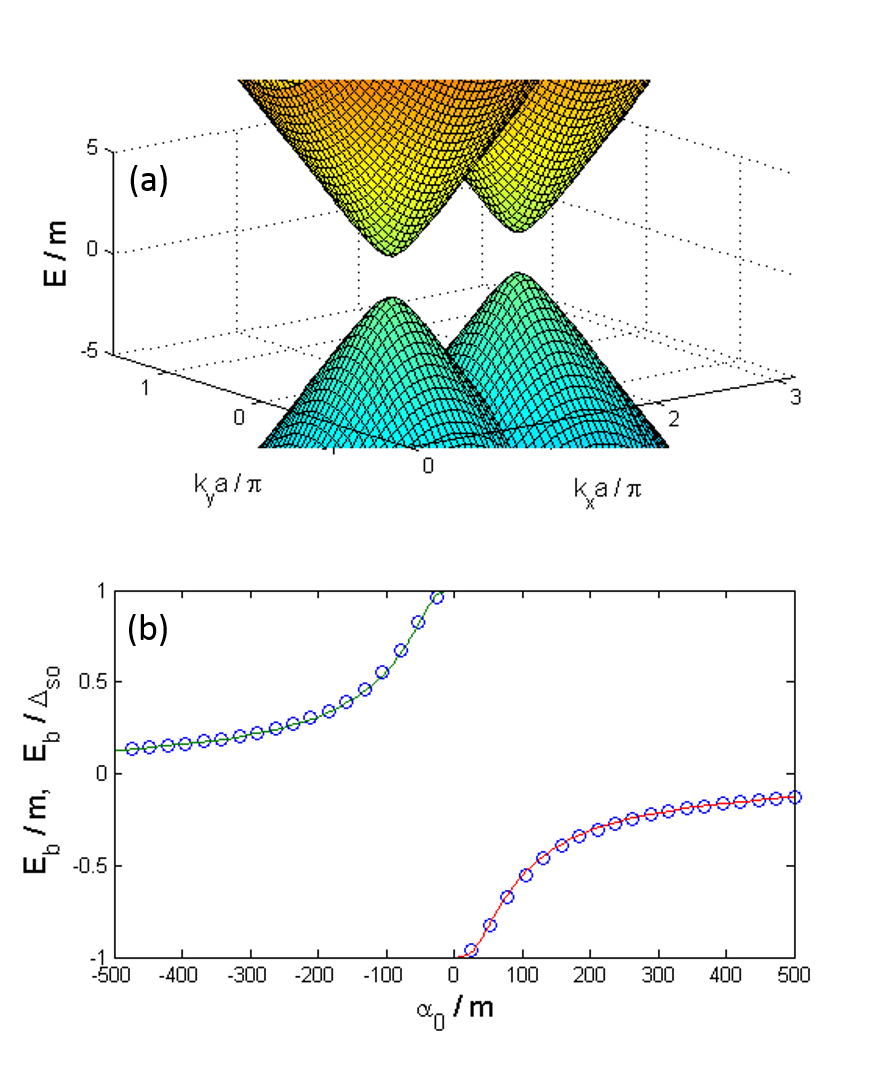}
\caption{ The Kane and Mele model is employed as an example of multiple minima band with $W/m=1.6$. The band gap appears near zero energy and its size is proportional to the spin-orbit coupling. The bound state energy associated with a single impurity is always within the gap, and they converge to zero energy at infinite impurity strength. The exact diagonalization result of the Kane-Mele model on the honeycomb lattice of $20 \times 20$ unit cells is overlaid for comparison with $\lambda_{SO} = m/6\sqrt{3}$ (open circles) so that the band gap of lattice model $\Delta_{SO}=6\sqrt{3}\lambda_{SO} = m$. The cutoff $W$ is chosen such that the dispersion relation of continuum model is a good approximation of that of the lattice model.  }
\label{fig:kanemele}
\end{figure}

As an example of an anisotropic system with multiple minima we consider an impurity problem in the Kane-Mele model\cite{kan2005}. This model describes electrons hopping on a honeycomb lattice with mirror-symmetric spin-orbit coupling.  It was the first model theorized to display a time reversal symmetric topological phase, i.e., the quantum spin Hall phase.  The two sublattices of the honeycomb lattice are encoded in two-dimensional spinors, and its band structure has two massive Dirac points in the Brillouin zone. For simplicity, but without loss of generality, we focus on a non-magnetic impurity and consider spinless fermions.  The Hamiltonian\cite{kan2005b} is:
\begin{eqnarray}
H=t\sum_{\left< ij \right>} c_i^\dagger c_j + i \lambda_{SO} \sum_{\left< \left< ij \right> \right>}\nu_{ij} c_i^\dagger s^z c_j.
\end{eqnarray}
 The first term is a nearest neighbor hopping, and the second term is spin-orbit interaction. $s^z$ is a Pauli matrix which acts in the spin space. $\nu_{ij}=(2/\sqrt{3})(\hat{d}_1\times \hat{d}_2)_z=\pm 1$, where $\hat{d}_1$ and $\hat{d}_2$ are unit vectors along the two bonds the electron traverses going from site $j$ to its next nearest neighbor $i$.  The effective Hamiltonian near the valley $K$ can be expressed by a $2\times2$ matrix in the pseudospin basis:
\begin{eqnarray}
\tilde{H}_K(q)&=&\begin{pmatrix} m& q e^{-i\theta} \\q e^{i\theta} & -m \end{pmatrix},
\end{eqnarray}
where $q$ is the momentum measured from $K$.
Since the linear spectrum is not well behaved in the presence of a $\delta$-function potential, we employ the Pauli-Villars regularization procedure as before. The regularized Hamiltonian reads:
\begin{eqnarray}
\tilde{H}'_K(q)&=& \tilde{H}_K(q)-\left[ E-\tilde{H}_K(q) \right] \frac{q^2}{W^2}.
\end{eqnarray}
The impurity is also described by a $2\times2$ potential matrix and we choose to put it on the $A$ site.
\begin{eqnarray}
\alpha =  \alpha_0  \begin{pmatrix} 1 & 0 \\ 0 & 0 \end{pmatrix}.
\end{eqnarray}
 To work this Hamiltonian in the sprit of 1-d wave function matching method, we ought to solve the real space 1-d Hamiltonian $H'_K(s)= \tilde{H}'_K(q \rightarrow \frac{\partial}{i\partial s} )$. Instead, let us work with real space Green function and then make use of Eq.\eqref{det} to find an associate bound energy of a single impurity. We follow the matrix version of Eq.\eqref{e1dsolution}: 
\begin{eqnarray}
G^{\theta}_K(s) &=& \int \frac{dk}{2\pi} \frac{e^{iks}}{E-\tilde{H}_K'(k)}, \\
&=&\sum_{j=1,2} \frac{ (-1)^j e^{-\lambda_j s}/2 }{|\lambda_j|(1+\frac{E^2-m^2}{W^2})}\begin{pmatrix} E+m& i \lambda_j e^{-i\theta} \\i\lambda_j e^{i\theta} & E-m \end{pmatrix},
\end{eqnarray}
where the poles of Green function are $\lambda_1=\sqrt{m^2-E^2}$, and $\lambda_2=W$ for $s>0$, and it is understood that $(-1)^{j=1}=-1$ and $(-1)^{j=2}=1$. The diagonal elements of the solution is symmetric of $s$, while the off diagonal elements are asymmetric. This is the wavefunction for the two sublattice system. For the other Dirac valley, K', the calculation is similar except the sign of the spin orbit coupling is opposite. Following Eq.\eqref{aniso} for anisotropic Hamiltonians:

\begin{eqnarray}
\frac{1}{\alpha_{K,K'}^{\theta}} = \frac{-1}{\pi}\int_0^{\infty} \frac{ds}{s}\frac{\partial G_{K,K'}^{\theta}(s)}{\partial s}.
\end{eqnarray}
The only remaining step is to consider the determinant from Eq. \eqref{det} and \eqref{det1}, in order to connect the potential strength and bound state energy:
\begin{eqnarray}
\rm{Det} \left[ I_2 +\sum_{m=K,K'}\int \frac{d\theta}{2\pi} \frac{1}{\alpha_{m}^{\theta}}\cdot{ \alpha } \right]=0.
\end{eqnarray}
As a result, we obtain the relation between the bound state energy and the potential strength $\alpha_0$ in the Kane-Mele model:
\begin{eqnarray}
\frac{1}{\alpha_0} = \frac{E}{2\pi^2(W^2-m^2+E^2)}\log\left( \frac{W}{\sqrt{m^2-E^2}} \right).
\end{eqnarray}
This is plotted in Fig.~\ref{fig:kanemele} in which we show both positive and negative impurity strengths and the associated positive and negative bound state energy. We compare our results for an exact diagonalization solution for the bound-state energy,

\section{Generalization to higher dimensions}\label{section:general}
 We have studied how to obtain the bound energy associated with a single impurity and found that the Green function at the impurity potential is a crucial quantity to find: 
\begin{eqnarray}
G(\vec{0} ) =\int \frac{d\Omega}{(2\pi)^{d-1}} \int^\infty _0 \frac{k^{d-1}dk}{2\pi} \frac{1}{E-\tilde{H}_d(\vec{k})}.
\label{Gd}
\end{eqnarray}
Although we have only discussed 2d systems, our method has a straightforward extension to higher dimensions. For odd dimensional cases, the wave function can be expressed by the summation of plane waves with certain momentum angles, since the k-integrals in Eq.\eqref{Gd} could be extended to the entire real k range. Then poles of the effective oriented Green function combine to produce the solution, and the "wave function matching" picture straightforwardly follows when relating to the impurity strength.

In this section let us consider a Hamiltonian in an even dimensional space. For simplicity we assume an isotropic Hamiltonian without any spinor structure. Then the bound state energy associated with the single impurity potential, $V(\vec{x}) = \alpha \delta^{2n}(\vec{x})$, is expressed similarly to Eq.\eqref{2dgeneral} by:
\begin{eqnarray}
\frac{1}{\alpha}
&=&  \int \frac{d^{2n}\vec{k}}{(2\pi)^{2n}} \frac{1}{E-\tilde{H}_{2n}(k)},\\
&=&\int \frac{k^{2n-1}dk}{2\pi} \frac{1}{E-\tilde{H}_{2n}(k)},
\end{eqnarray}
where the difficulty lies in the integration in polar coordinates; as in the 2d case,  the contour integration is not readily possible, and therefore any connection to a semi-classical beam analysis is not possible.   Nevertheless, to overcome this hurdle we may use the same prescription as in 2d.  We use the Kramers-Kronig relation to make use of the symmetric imaginary part of integrand and obtain the generalized relation:
\begin{eqnarray}
\frac{1}{\alpha}
&=&  \frac{-1}{\pi} \int_{0}^{\infty} \frac{ds}{s} \frac{\partial^{2n-1}}{\partial s^{2n-1}} \left[ \int_{-\infty}^{\infty} \frac{dk}{2\pi} \frac{e^{iks}}{E-\tilde{H}_{2n}(k)} \right].
\end{eqnarray}
For isotropic and single-minimum band type problems, the above relation can be translated to finding an impurity state in a renormalized-strength impurity potential in the 1-d effective Schr\"odinger equation:
\begin{eqnarray}
\left[E-H_{1d}(\partial_x) \right]\psi_{1d} (x) =  \delta (x) \left[ \frac{-\alpha}{\pi}\int_{0}^{\infty} \frac{ds}{ s} \frac{\partial^{2n-1}}{\partial s^{2n-1}}\psi_{1d} (s) \right],
\end{eqnarray}
where  $H_{1d}(\partial_x) = \tilde{H}_{2n}(k \rightarrow \frac{\partial}{ i\partial x})$ , and the solution is the sum of plane waves with complex wave numbers:
\begin{eqnarray}
\psi_{1d}(x) =\int_{-\infty}^{\infty} \frac{dk}{2\pi} \frac{e^{iks}}{E-\tilde{H}_{2n}(k)}.
\end{eqnarray}

\section{Conclusions}

Finding impurity bound states in 1d is quite intuitive, and is carried out by combining plane-wave states into a consistent solution.  The plane-wave approach to solving impurity problems in 1d arises naturally when considering the T-matrix approach, or, equivalently, the Green function expression of Eq.\eqref{18}. The contour integration of the 1d Green function results in a discrete sum of pole-contributions, each of which is associated with plane-wave eigenstate of the uniform Hamiltonian. In 1d, the Schr\"odinger equation can be used directly to obtain impurity states, and such a calculation would involve satisfying matching conditions at the impurity location of plane-wave solutions - the same that arise from the poles of the Green function -  belonging to either side. The intuitive interpretation of impurity states as a simple combination of plane-waves is completely lost at higher (even) dimensions, which is seen technically by not being able to reduce Eq.\eqref{18} to the sum of residues of the Green function.

In this manuscript we presented a new approach for finding bound states, which reduces the impurity problem in any dimension to 1d impurity problems, and thus allows an interpretation in terms of a small set of incoming and outgoing plane waves on a linear trajectory.

We demonstrated that the method could be efficiently used to find impurity bound states in a general band structure where the gap could have multiple minima, and a spinor structure. Our method relied on the use of the Kramers-Kronig relation, which maps the Green-function formula at any dimension, to an expression which is given again by a sum of residues corresponding to plane-wave solutions of the pure model.

Presenting a few examples, we demonstrated how our method easily lends itself to approximating lattice Hamiltonians in terms of a discrete sum of separate valley Hamiltonians. As we show, when Pauli-Villars regulators are used to provide a cut-off for valley Hamiltonians, we can still use the Kramers-Kronig relation to connect bound states with a discrete sum of plane waves. The Pauli-Villars regulators, however, add additional poles to the Green function of the pure system, which also need to be included in the plane-wave superposition.

While we only demonstrated the method in 2d, the wavefunction matching method can be extended to any dimension. All odd dimensional systems are analogous to the 1d case with additional angular variables, and all even dimensional systems are analogous to the 2d case. This is because the wavefunction matching method is closely connected to the contour integration which reduces the problem to finding poles of Green function; the contour integration applies in odd dimensions but not in even ones. For most systems  without  rotational symmetry of local Hamiltonians, only a few momentum angles are necessary to complete the semiclassical interpretation of a bound state.

KWK and GR would like to acknowledge support from DARPA through FENA, as well as from the IQIM, an NSF center with the support of the Gordon and Betty Moore Foundation. TPB would like to acknowledge support from NSERC and FQRNT.

\bibliographystyle{apsrev}
\bibliography{semi_ref}

\end{document}